\documentclass[preprint,3p,authoryear]{elsarticle}
\journal{Icarus}

\pdfoutput=1

\usepackage{natbib}

\usepackage{amssymb}
\usepackage{amsmath}

\usepackage{lineno}
\usepackage[title]{appendix}
\usepackage{textcomp}
\usepackage[version=4]{mhchem}
\usepackage{color}
\usepackage{ulem}
\usepackage{doi}

\usepackage{hyperref}
\hypersetup{colorlinks=true, urlcolor=blue}

\begin{document}

\begin{frontmatter}



\title{Close-up images of the final Philae landing site on comet 67P/Churyumov-Gerasimenko acquired by the ROLIS camera\tnoteref{label1}\tnoteref{label2}}
\tnotetext[label1]{\doi{10.1016/j.icarus.2016.12.009}}
\tnotetext[label2]{\copyright 2017. This manuscript version is made available under the CC-BY-NC-ND 4.0 licence\\ \url{https://creativecommons.org/licenses/by-nc-nd/4.0/}}


\author[DLR]{S.E.~Schr\"oder\corref{cor1}}
\ead{stefanus.schroeder@dlr.de}
\author[DLR]{S.~Mottola}
\author[DLR]{G.~Arnold}
\author[DLRb]{H.-G.~Grothues}
\author[DLR]{R.~Jaumann}
\author[DLR,TUB]{H.U.~Keller}
\author[DLR]{H.~Michaelis}
\author[IAS]{J.-P.~Bibring}
\author[DLR]{I.~Pelivan}
\author[DLR]{A.~Koncz}
\author[DLR]{K.~Otto}
\author[CNES]{E.~Remetean}
\author[Mag]{F.~Souvannavong}
\author[Mag]{B.~Dolives}

\cortext[cor1]{Corresponding author.}

\address[DLR]{Deutsches Zentrum f\"ur Luft- und Raumfahrt (DLR), 12489 Berlin, Germany}
\address[DLRb]{Deutsches Zentrum f\"ur Luft- und Raumfahrt (DLR), 53227 Bonn, Germany}
\address[TUB]{Institut f\"ur Geophysik und extraterrestrische Physik (IGEP), Technische Universit\"at Braunschweig, 38106 Braunschweig, Germany}
\address[IAS]{Institut d'Astrophysique Spatiale (IAS), 91405 Orsay, France}
\address[CNES]{Centre National d'\'Etudes Spatiales (CNES), 31400 Toulouse, France}
\address[Mag]{Magellium, 31521 Ramonville Saint-Agne, France}

\begin{abstract}

After coming to rest on the night side of comet 67P/Churyumov-Gerasimenko, the ROLIS camera on-board Rosetta's Philae lander acquired five images of the surface below the lander, four of which were with the aid of LED illumination of different colors. The images confirm that Philae was perched on a sloped surface. A local horizon is visible in one corner of the image, beyond which we can see the coma. Having spent a full day on the surface Philae was commanded to lift and rotate, after which a final, sixth, LED image was acquired. The change in perspective allowed us to construct a shape model of the surface. The distance to the foreground was about 80~cm, much larger than the nominal 30~cm. This caused stray light, rather than directly reflected LED light, to dominate the image signal, complicating the analysis. The images show a lumpy surface with a roughness of apparently fractal nature. Its appearance is completely different from that of the first landing site, which was characterized by centimeter to meter-sized debris \citep{M15}. We recognize neither particles nor pores at the image resolution of 0.8~mm per pixel and large color variations are absent. The surface has a bi-modal brightness distribution that can be interpreted in terms of the degree of consolidation, a hypothesis that we support with experimental evidence. We propose the surface below the lander to consist of smooth, cracked plates with unconsolidated edges, similar to terrain seen in CIVA images.

\end{abstract}


\end{frontmatter}

%\begin{linenumbers}

\section{Introduction}
\label{sec:introduction}

Activity on 67P/Churyumov-Gerasimenko is primarily governed by solar insolation \citep{K15}. Its surface probably consists of a desiccated, dust-rich layer that blankets an interior enriched in water ice, where sublimation takes place as soon as the insolation conditions are favorable. However, how cometary activity works on small scales is largely unknown \citep{G15}. Does gas permeate through micropores in the active surface, or does outgassing occur from discrete macroscopic features like cracks and vents? Is outgassing a continuous phenomenon or does it occur in discrete bursts? The structure of the upper surface layer (sub-mm to dm) is probably the key regulator of cometary activity. This layer controls the thermal balance, the heat transport to the interior, and the gas transport to the exterior. It is heavily shaped and eroded by the activity itself; fragments are regularly ejected and deposited \citep{T15}. Yet, cometary activity is maintained, apparition after apparition. It was one of the main science goals of the Philae lander on the Rosetta spacecraft to understand the role of the upper surface layer in cometary activity. The high-resolution images of the Rosetta lander imaging system (ROLIS) acquired after landing contribute to the first-ever close-up characterization of the upper layer. ROLIS is a highly miniaturized CCD camera looking straight down from its vantage point on the instrument balcony \citep{M07}. During Philae's initial descent towards the comet surface, ROLIS successfully acquired images of the comet surface at the Agilkia site \citep{M15}. After bouncing off the surface, Philae interacted with the surface several times more over a period of two hours, before arriving at its final landing site, Abydos, which is located about 1~km away from Agilkia \citep{B15}. Philae has now been identified in images of the Rosetta OSIRIS camera at the location proposed by \citet{C15}. The images show the lander wedged between what appear to be meter-sized rubble fragments in the shadow of a large outcrop\footnote{\url{http://www.esa.int/Our_Activities/Space_Science/Rosetta/Philae_found}}.

\section{ROLIS camera}

ROLIS is a dual-purpose camera, designed to work both as a descent imager and as a close-up imager after touch-down. The technical details of the experiment are described by \citet{M07}, and we provide a short summary here. The camera was able to switch from the former to the latter mode by refocusing to a nominal object distance of 30~cm by shifting the foremost optical element, the Infinity Front Lens (IFL), out of the optical path. The field-of-view (FOV) is $57.7^\circ \times 57.7^\circ$. The images acquired by the frame transfer CCD detector were transmitted with a resolution of 14~bits, i.e.\ a maximum of 16383 data numbers (DN). ROLIS made use of four LED arrays to illuminate the near-field scene below the lander in the following colors: {\it Blue}, {\it Green}, {\it Red}, and {\it IR}, spanning the  wavelength range from 400 to 950~nm. The LED names reflect their colors as perceived by the eye, with the {\it IR} flux curve peaking at 880~nm. The camera had been pre-programmed to take a set of five images after landing, one image for each LED color and one image ({\it Dark}) with the LEDs switched off. The latter image was expected to be truly dark exposure, as imaging was scheduled to start shortly after local midnight. The images are listed in Table~\ref{tab:images} in order of acquisition and are referred to by their LED color or sequential number (e.g.\ the {\it Blue} image is \#1). The full post-landing imaging sequence was successfully executed and all images were returned. An auto-exposure algorithm was implemented to ensure correct illumination levels, given the a-priori uncertainty about the distance to the scene (i.e.\ the intensity of reflected light) and the image contrast. This simple algorithm was based on a histogram analysis of test images, with the final decision on the exposure time controlled by these user-defined, commandable parameters: the DN level and maximum percentage of overexposed and underexposed pixels, respectively. First ROLIS removed the IFL from the optical path and then acquired following images in this order: {\it Blue}, {\it Green}, {\it Dark}, {\it Red}, and {\it IR}. After two (Earth) days on the surface, during which the Philae body was slightly lifted and rotated to re-orient its solar panels, the opportunity arose to take a final image. A second {\it Red} image was acquired with the IFL in the optical path. In a final dramatic act, Philae transmitted this last image to Rosetta only seconds before its battery power ran out.

The ROLIS CCD has an active area of $1024 \times 1024$ pixels that is surrounded by several columns and rows that are covered by an opaque mask. For the post-landing images the parts of the image containing the covered regions were discarded on-board and only the active area was transmitted. An exception was made for the first image transmitted: a thin strip sized $1066 \times 30$ pixels that was acquired without illumination and essentially shows only readout noise (not listed in Table~\ref{tab:images}). We calibrated the raw images as follows. First we subtracted a bias value of 214~DN from the entire image. At a camera temperature of $-72^\circ$C, dark current was essentially zero and a correction is not necessary. The rapid shift of the image from the CCD active area into a storage area introduces a smear over the image, but since the shift time (3.2768~msec) is so much smaller than the exposure times of the post-landing images (see Table~\ref{tab:images}), a correction is not necessary. For example, for the worst case of image~\#4 the smear is expected to contribute a top-to-bottom gradient of a few DN. We divided the images by a flat field exposure acquired before launch. This flat field was acquired with the IFL in the optical path, and it exhibits large rings of stray light (more about this below). The use of this flat field must have introduced these rings into the calibrated images \#1, \#2, \#4 and \#5 (which were acquired without the IFL), but so subtly that we have not been able to find clear evidence for that. The last step in the calibration was division by the exposure time. The result is the intensity in units of DN per second, albeit flat-field-corrected. We cannot convert this intensity to surface reflectance units because the degree of aging and contamination of the optics and the LED arrays are unknown, and neither do we know the distance to the illuminated surface with sufficient accuracy. Due to the limited data link budget of Philae, it was necessary to apply lossy compression to all the images to reduce the data volume. The compression algorithm is based on wavelet integer decomposition and zero-tree encoding. The compression level was set to achieve a compression factor of about 8. All images are archived in the Planetary Data System\footnote{\url{https://pds.nasa.gov}} and the Planetary Science Archive\footnote{\url{http://www.cosmos.esa.int/web/psa/psa-introduction}}, both in raw form and calibrated as described here.

\section{Image analysis}

\subsection{Image overview}
\label{sec:overview}

\begin{figure}
\centering
\includegraphics[width=\textwidth]{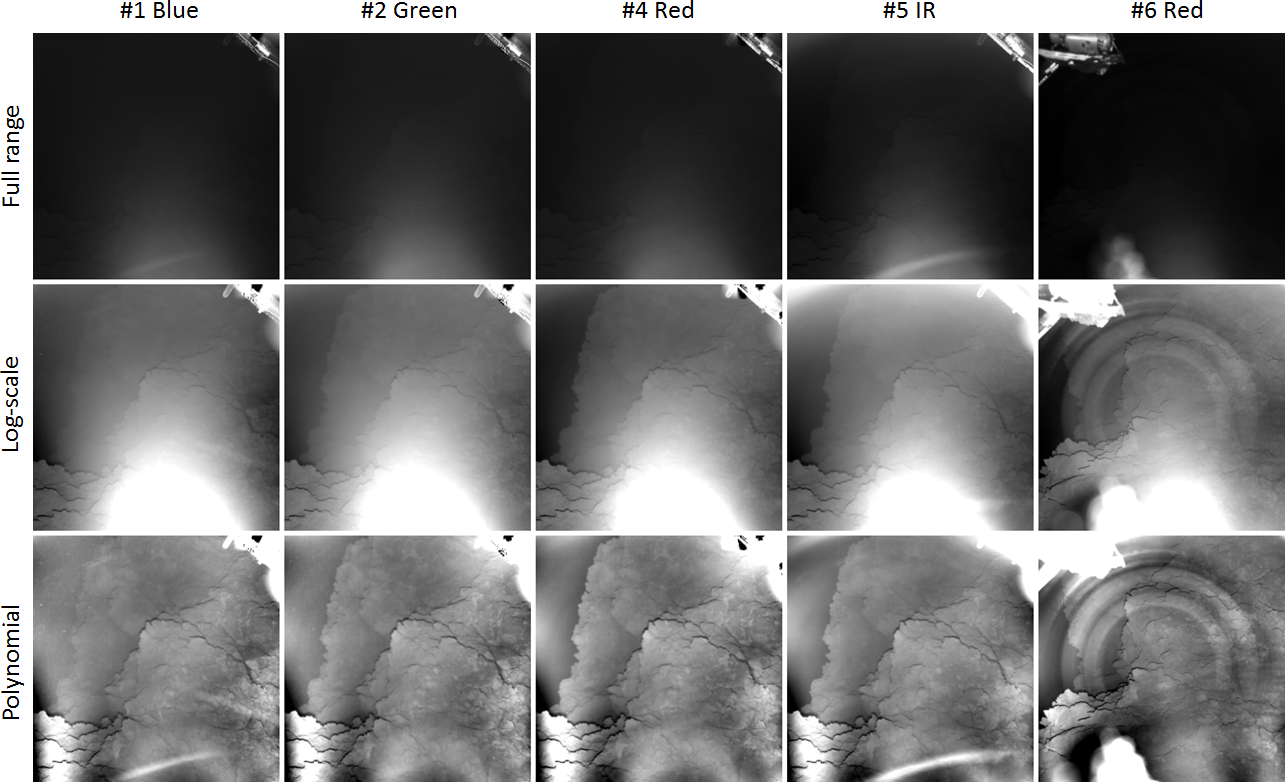}
\caption{Overview of all LED images acquired after landing. {\bf Top} row: linear scale brightness, full dynamic range. {\bf Middle} row: log-scale brightness, clipped. {\bf Bottom} row: images divided by a best-fit polynomial surface. Note that the last method enhances surface features but does not preserve relative brightness levels. Image numbers refer to entries in Table~\ref{tab:images}.}
\label{fig:image_overview}
\end{figure}

All images acquired with LED illumination are shown in Fig.~\ref{fig:image_overview}. At their full dynamic range (top row) the images appear mostly dark. A partly overexposed part of the Philae $+X$ lander leg can be seen in the top right corner of all images but the last, and the comet surface is visible at the bottom. In the last {\it Red} image the Philae harpoon \citep{B15} is visible in the upper left corner, and at the bottom an out-of-focus object appears to be dangling in the field-of-view (FOV), that we suspect to be a cable of the MUPUS experiment \citep{S07}. The comet surface is actually well exposed in all images, but appears dark in this representation because of the presence of bright lander objects in the FOV. The peak surface brightness is at the bottom of the images instead of the center and there is a brightness gradient from top to bottom, which implies that the surface is strongly inclined with respect to the lander. To see the surface more clearly we enhance the images. We do this in two ways. First, we display the image brightness on a logarithmic scale and clip it to such limits that most of the surface becomes visible (Fig.~\ref{fig:image_overview}, middle row). This artificially saturates the spacecraft structures and the surface at the bottom on the image. The surface in images \#1-\#5 is slightly out-of-focus. As the camera was designed to be well focused for a nominal landing, this implies that the surface was farther away than 30~cm. The last image is fully in focus as the IFL had been moved in the optical path. However, it also features subtle concentric rings of stray light associated with the IFL, which become apparent after enhancement. These rings are also present in two ROLIS images acquired directly after Philae's separation from Rosetta. Even though the flat field used for calibration also has these rings of stray light, division by the flat field does not suppress them in the comet image because stray light must be subtracted. An alternative way to enhance the surface features is dividing the images by a fourth-order polynomial surface fitted to selected parts of the individual images (Fig.~\ref{fig:image_overview}, bottom row). This method reveals subtle local surface brightness variations, even though it does not preserve brightness variations over large scales. For example, the brightness of the surface in the lower left corner is over-enhanced. For clarity, we show the (enhanced) last image acquired at its full resolution in Fig.~\ref{fig:last_red}. Details on the polynomial surfaces are provided in the supplementary material to this paper.

\begin{figure}
\centering
\includegraphics[width=8cm]{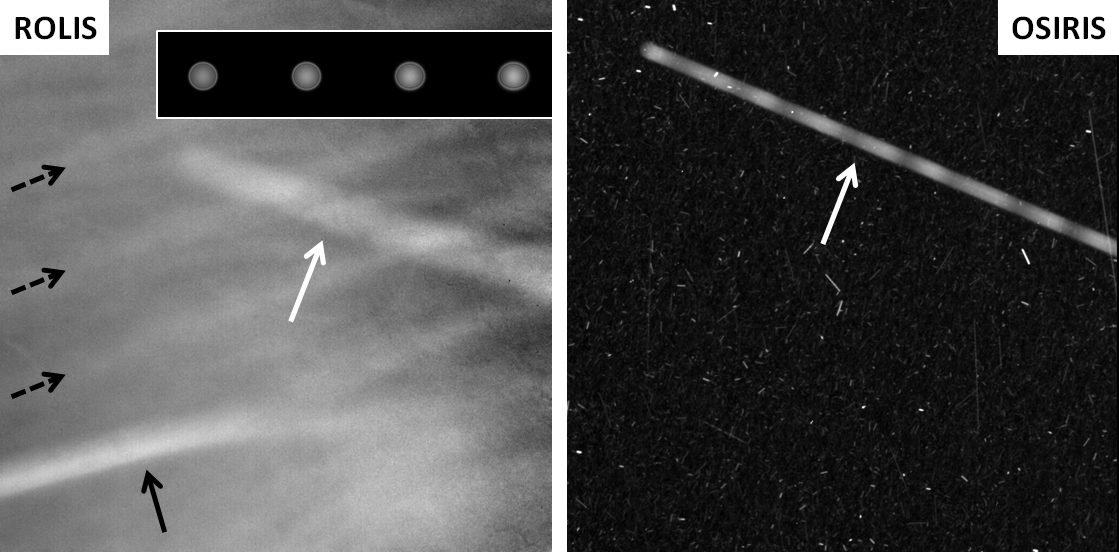}
\caption{Suspected particles (white arrows) traveling through the FOV during the image exposure. {\bf Left}: Ratio of the {\it Blue} and {\it Green} images (bottom right quadrangle). The arch-like feature (black solid arrow) and the diagonal pattern (dashed arrows) are associated with defects in the camera optics. The inset shows the PSF for a point source at 6~cm distance along the particle track. The track is slightly wider than the PSF size, suggesting the particle was closer to the camera than 6~cm. {\bf Right}: Rotating dust grain in the coma seen by OSIRIS NAC on 4 March 2016 (credits: ESA\slash Rosetta\slash MPS for OSIRIS Team MPS\slash UPD\slash LAM\slash IAA\slash SSO\slash INTA\slash UPM\slash DASP\slash IDA).}
\label{fig:particle}
\end{figure}

The comet surface that becomes visible after enhancement looks completely different from that at the first landing site \citep{M15}. Illumination by the LEDs contributes to the surface's unusual appearance. The small distance between camera aperture and LEDs leads to a very small phase angle over the whole image, causing reduced shading and thin, narrow shadows that create the impression of some form of organization, perhaps layering. The surface shows abundant small scale structure, but individual particles that make up the surface cannot be distinguished. At first glance it is impossible to tell whether we are looking at a hard or a soft (fluffy) surface. Larger particles like cobbles and pebbles are absent, even though they were abundant at the first landing site \citep{M15}. Furthermore, the images have a distinct hazy appearance, meaning a reduced contrast at larger distance, which is unexpected for conditions of vacuum. The area in the upper left corner seems to show no surface features but has an otherwise ``dirty'' appearance with tiny smudges and specks. As we explain below, our assessment is that here ROLIS was peering over a local horizon towards the coma. Enhancing the images reveals more puzzling features. One of these is a broad streak in the bottom right corner of the {\it Blue} image (Fig.~\ref{fig:particle}). We interpret this streak as an out-of-focus particle traveling past the camera at a distance much closer than 30~cm. It is slightly tapered, which implies that the particle moved either towards or away from the camera. A simulation of the point spread function (PSF) suggests the particle passed by at a distance of about 4-5~cm (Fig.~\ref{fig:particle}, inset). Such streaks are not seen in the other ROLIS images, but similar ones have been observed by the OSIRIS camera on-board Rosetta and assumed to be due to particles passing close to the camera \citep{Fu15}.

Other spurious features are the diagonal pattern of lines in the lower half of all ROLIS images and a bright arch at the bottom of the {\it Blue} and {\it IR} images (but not the {\it Green} and {\it Red} images). The location of the lines is slightly different for each color, creating the impression of rainbow-like fringes that suggest an interference origin (see Sec.~\ref{sec:variations}). Both the lines and the arch are seen in images of the inside of Rosetta acquired by ROLIS during cruise, and we believe that they result from damage to the camera optics incurred after spacecraft integration, possibly during launch. A smaller feature that looks like a scratch (indicated in Fig.~\ref{fig:coma}) is also only present in the {\it Blue} and {\it IR} images and is most likely also an optical artifact.

\begin{figure}
\centering
\includegraphics[width=\textwidth]{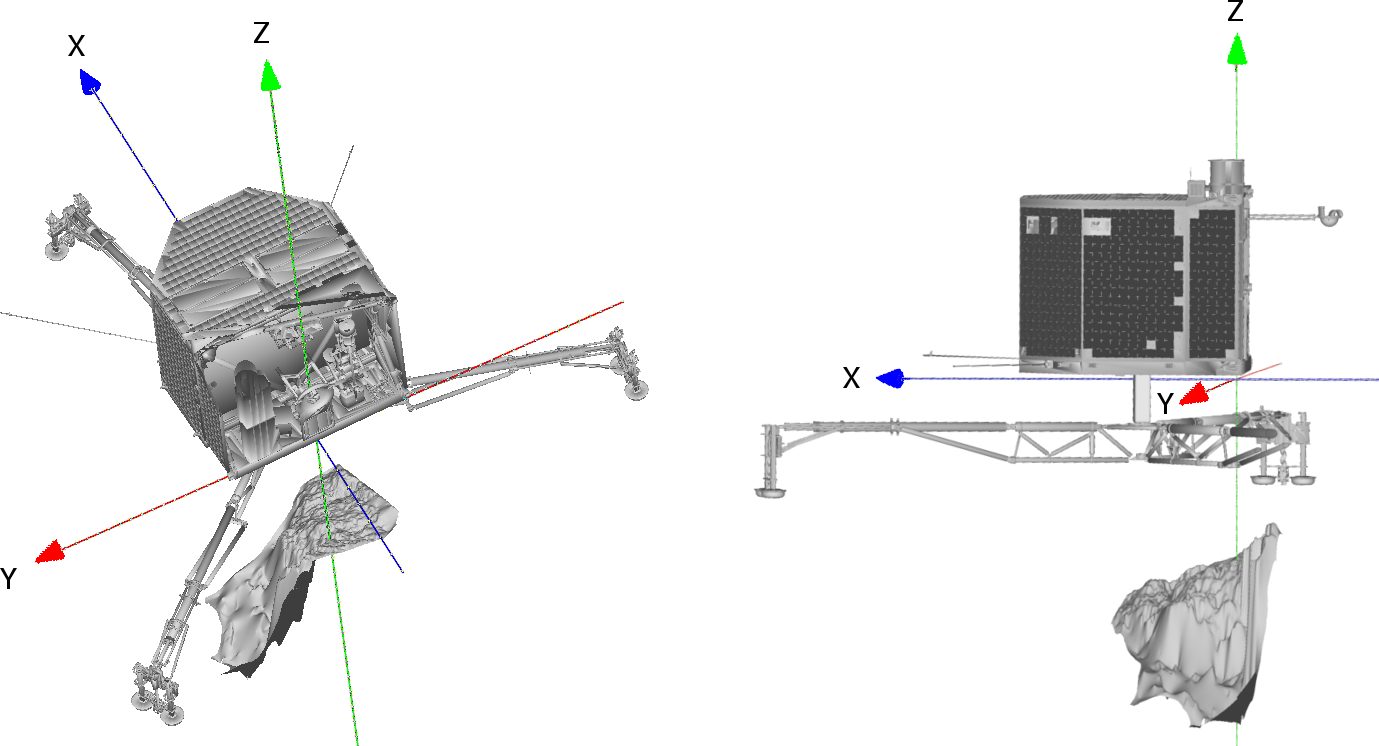}
\caption{Perspective views of the digital terrain model (DTM) of the surface seen in the ROLIS images, shown for two orientations of the Philae lander in its final position. The axes define the ROLIS reference system with the aperture located at the origin. The ROLIS and Philae reference systems have the same orientation but a slightly different origin.}
\label{fig:DTM}
\end{figure}

\begin{figure}
\centering
\includegraphics[width=\textwidth]{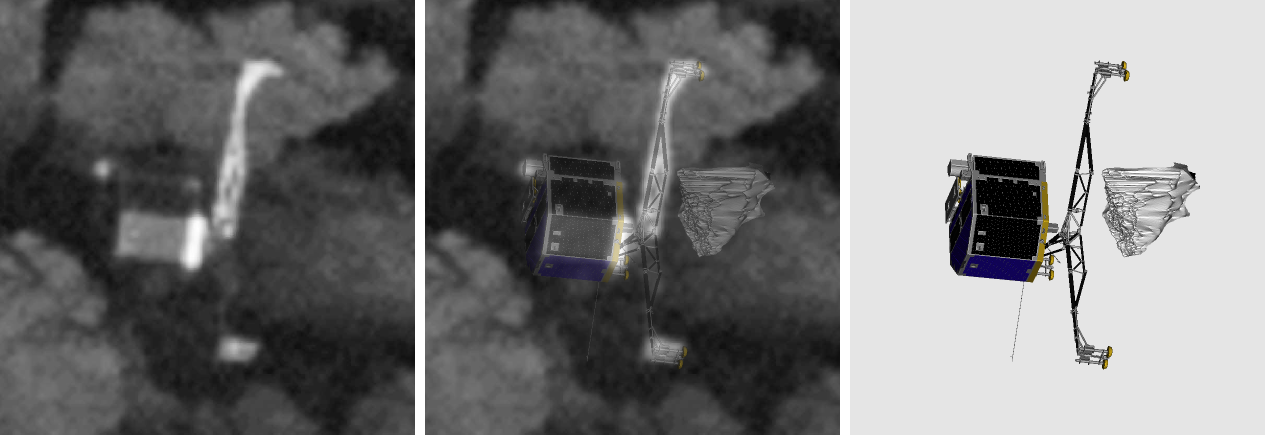}
\caption{Philae on the comet surface as seen by the OSIRIS camera ({\bf left}) with a transparent Philae model superposed ({\bf center}). The image at {\bf right} shows only the Philae model, including the surface DTM from Fig.~\ref{fig:DTM}. OSIRIS image credits: ESA\slash Rosetta\slash MPS for OSIRIS Team MPS\slash UPD\slash LAM\slash IAA\slash SSO\slash INTA\slash UPM\slash DASP\slash IDA.}
\label{fig:Philae_found}
\end{figure}

Between acquisition of the {\it IR} (\#5) and the second {\it Red} (\#6) images the lander body was commanded to lift with respect to the surface by about 10~cm and rotate clockwise by about $22^\circ$. The change in perspective allowed us to reconstruct a digital terrain model (DTM) from the ROLIS images with the method described by \citet{G05}. The DTM in Fig.~\ref{fig:DTM} shows that the surface in the lower half of the images is in the foreground, at a distance of 0.7-0.9~m. Given the FOV, this amounts to a spatial resolution of about 0.8~mm per pixel. Towards the local horizon the surface slopes away steeply, and the horizon itself is about 1.5~m away. We estimate the angle between the ROLIS boresight vector and the normal of the sloping surface (the inclination) to be about $60^\circ$. Philae has now been located on the comet surface by the OSIRIS camera. The image\footnote{\url{http://www.esa.int/spaceinimages/Images/2016/09/Philae_found}} in Fig.~\ref{fig:Philae_found} shows the lander lying completely on its side. This orientation is fully consistent with the puzzling aspects of the ROLIS images described above, and may have allowed parts of other instruments to invade the FOV. We reconstructed Philae's orientation by fitting a model of the lander to the OSIRIS image (Fig.~\ref{fig:Philae_found}). Given the small apparent size of the lander, our reconstruction reproduces the image well. Figure~\ref{fig:Philae_found} also shows the correct relative orientation of the surface DTM from Fig.~\ref{fig:DTM}. Even though the signal-to-noise (S/N) of the surface below the lander in this particular OSIRIS image is relatively low, its shape seems to agree well with the DTM.

\begin{figure}
\centering
\includegraphics[width=8cm]{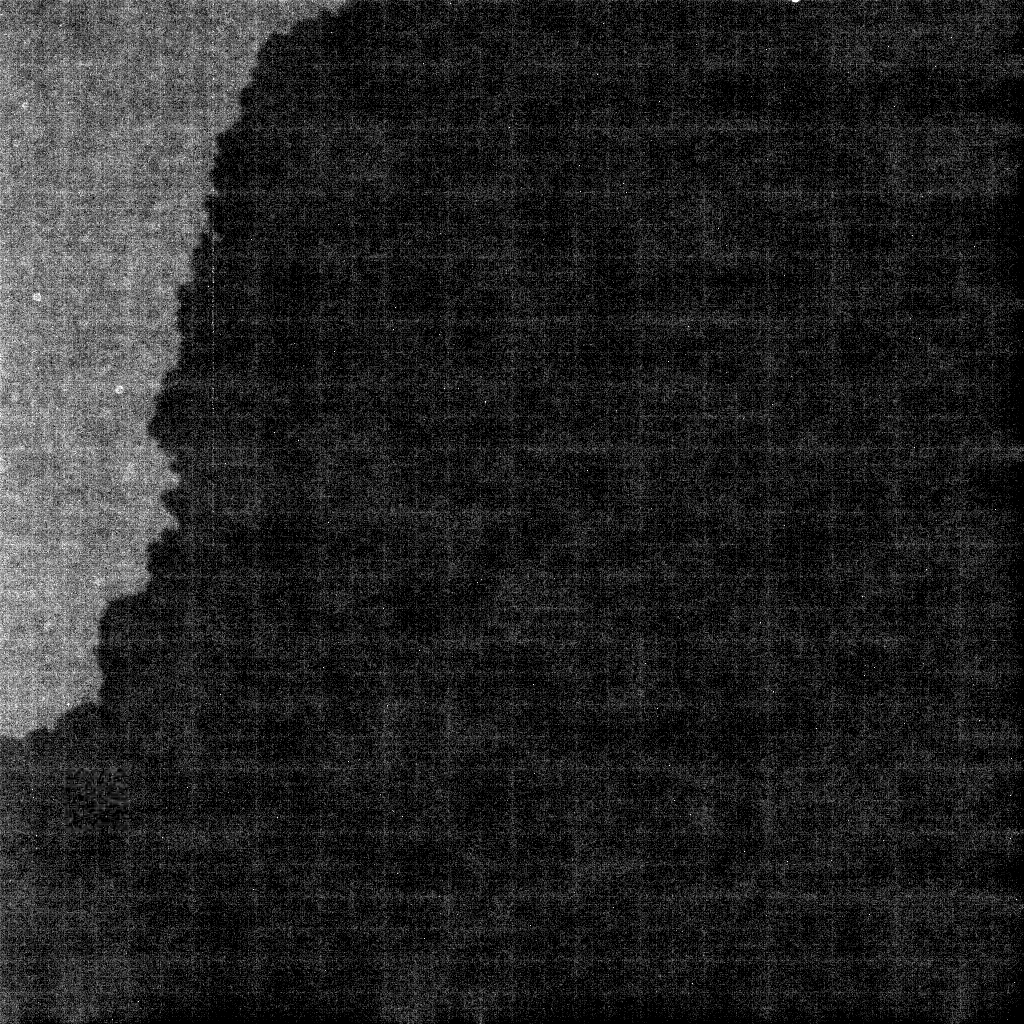}
\caption{{\it Dark} image \#4 (Table~\ref{tab:images}) acquired without LED illumination, with the brightness scaled between 215~DN (black) and 230~DN (white). The square pattern is due to the lossy compression scheme.}
\label{fig:dark}
\end{figure}

Between the {\it Green} (\#2) and {\it Red} (\#4) images ROLIS acquired the {\it Dark} exposure with the LEDs off. The resulting image (\#3, Fig.~\ref{fig:dark}) was expected to be uniformly dark at the bias level of about 215~DN, as the camera temperature was too low for dark current to be measurable. However, while most of the image is indeed dark, the upper left corner of the image is about 15~DN brighter and separated from the dark part of the image by a jagged contour. Other than that the image shows a faint square pattern of compression artifacts. Dark images acquired during cruise are uniformly dark and only show compression artifacts.

\subsection{Coma}

\begin{figure}
\centering
\includegraphics[width=\textwidth]{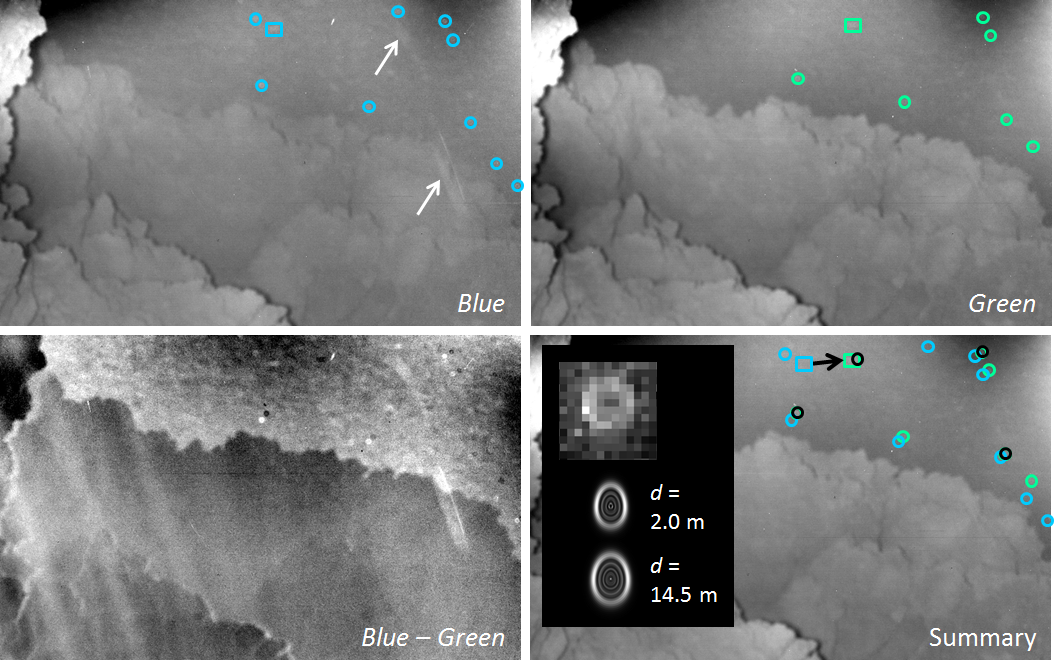}
\caption{Distant particles moved in the coma visible beyond the local horizon. They can be tracked in three consecutive images ({\it Blue}, {\it Green}, and the dark). Each particle detection is indicated by a circle, and the rectangles in the {\it Blue} and {\it Green} images denote particle trails. The arrows in the {\it Blue} image point at instrumental artifacts. The difference between the {\it Blue} and {\it Green} frames is shown at bottom left. The summary frame at the bottom right shows how the locations line up, with the blue, green, and black symbols indicating the particle position in the {\it Blue}, {\it Green}, and dark images, respectively. The inset compares one particle image (enlarged) with two simulations of the PSF of a point source located at distances $d$ of 2.0 and 14.5 m from the camera, calculated for the location in the image frame of the particle image. Note that the particle image is affected by motion smear of a few pixels.}
\label{fig:coma}
\end{figure}

The jagged contour in the {\it Dark} image (Fig.~\ref{fig:dark}) represents a local horizon, and the region of elevated brightness is the Sun-illuminated coma beyond, which confirms that Philae was strongly inclined (Fig.~\ref{fig:Philae_found}). The coma region is not empty but has a rather ``dirty'' appearance. Small ring-like blobs can be distinguished in the dark image and the enhanced {\it Blue} and {\it Green} images (Fig.~\ref{fig:coma}). Upon closer inspection these ringlets can be traced from one image to the other and are found to travel on linear trajectories. No such ringlets can be found below the horizon and therefore must be restricted to the coma. The ringlets are seen in the {\it Blue} and {\it Green} images but not the {\it Red} and {\it IR} images. The difference between these two groups is the exposure time, which is long for {\it Blue} and {\it Green} (25-30~sec) and short for {\it Red} and {\it IR} (5-6~sec). If the ringlets represent particles that are illuminated by the LEDs they should be present in all images but the dark one. We conclude that these are Sun-illuminated particles traveling through the coma. Their ring-like appearance is consistent with PSF simulations of distant point sources (Fig.~\ref{fig:coma}, inset). One of these particles (indicated with an arrow in the summary frame of Fig.~\ref{fig:coma}) traveled so fast with respect to Philae that it displays a track (motion blur). The track lengths in the {\it Blue} and {\it Green} images are consistent with the separation between the tracks when assuming a constant particle velocity. While we can conclude from the simulations that the particles are at least several meters away, consistent with a location in the coma, our analysis is unfortunately by far not sensitive enough to establish a precise distance. Thus we are also not able to provide estimates of the particle velocities. The particles are all moving in the same general direction, which may represent a direction away from the comet. Our interpretation is consistent with that of \citet{Bib15}, who also interpreted a bright spot observed by CIVA camera~1 as a detached grain floating around the nucleus. The coma was awash with particles around the time of Philae's landing (e.g.\ \citealt{Fu15}).

\subsection{Stray light}
\label{sec:stray_light}

\begin{figure}
\centering
\includegraphics[width=\textwidth]{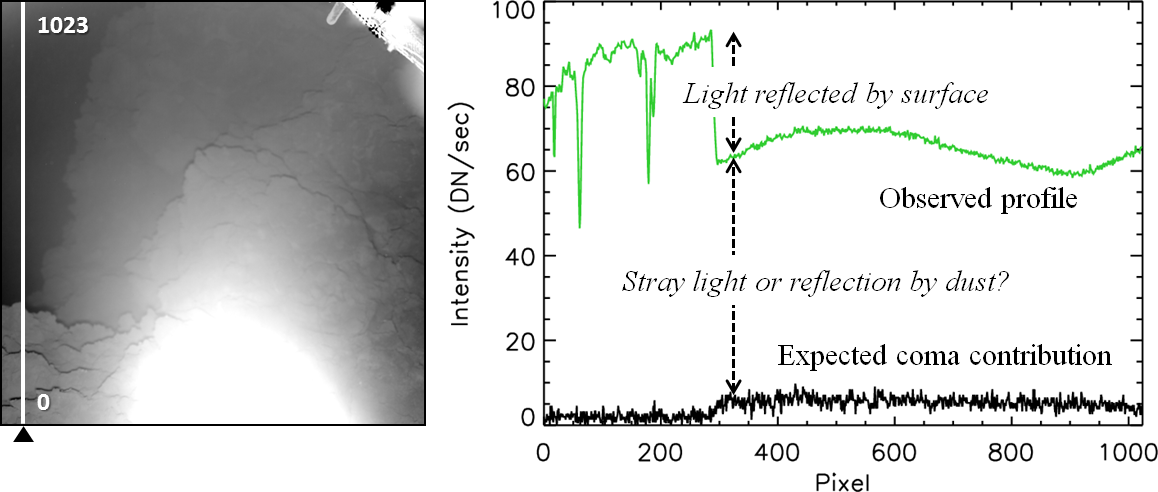}
\caption{All images have significant signal in the ``empty'' space beyond the horizon. From the dark image we can calculate the expected signal from the coma. The difference must be attributed to either stray light or LED light scattered by suspended very small particles close to the camera. In this case, the light reflected by the surface makes up only a third of the total signal. The profiles shown here are column~50 of the {\it Green} image (green line) and the dark image (black line). Pixel numbers are indicated in the left figure.}
\label{fig:stray_light}
\end{figure}

The ROLIS images in Fig.~\ref{fig:image_overview} have a distinctly ``foggy'' appearance: the contrast appears to be reduced in some parts of the frame and shadows cast by the LED illumination are not completely black. There is also significant signal in the ``empty'' region beyond the horizon. We can predict the coma signal in this region from the dark image, and the observed signal is larger than predicted by one to two orders of magnitude, depending on LED color and location within the image (Fig.~\ref{fig:stray_light}). The largest excess signal we find in the {\it IR} image and the smallest in the {\it Blue} and {\it Green} images. The extra signal might be due to LED light scattered off very small (because unresolved) particles suspended close to the camera. An alternative explanation is the signal being due to stray light off the lander leg in the upper right corner and other lander structures outside the FOV. If the excess signal were exclusively due to Rayleigh scattering by smoke-sized particles, it should be largest for {\it Blue}, which is opposite to what we observe. It could also be due to the presence of larger, but still unresolved, suspended particles (with diameters in the range of about 1-100~\textmu m). The spectral dependence of light scattered by such particles is determined by the distribution of particle shape, size, and refractive index, and therefore unknown. Stray light is a viable alternative. It was not possible to characterize the stray light characteristics of the camera flight model before launch. We did, however, perform stray light experiments with the flight spare using a white target. The results confirmed the existence of in-field stray light, which was strongest and most pronounced in the {\it IR} image and weakest in the {\it Blue} and {\it Green} images. Considering {\it IR} image \#5, we note that the excess signal in the coma region is largest of all colors and both the top corners are distinctly brighter. In the top right corner is a lander leg and in the top left corner, just outside the FOV, is another Philae structure (it is visible in the descent images). We conclude that the {\it IR} image is definitely affected by stray light. Stray light is probably pervasive in all images over the entire FOV. Also, the bottom center of the images may have been substantially brightened by stray light from an unknown source outside the FOV, most likely a part of another instrument, a suspicion fed by the appearance of the experimental surfaces discussed later in this paper (Sec.~\ref{sec:experiment}). The contribution of light reflected by the comet surface to the observed signal is small (see Fig.~\ref{fig:stray_light}) because the surface was much farther away than anticipated. This reduced the LED flux arriving at the comet surface by more than an order of magnitude.

Unfortunately, we cannot correct the images for stray light. The exact stray light distribution over an image depends on the position and reflective properties of the stray light sources, the angular distribution of the LED flux, and the stray light characteristics of the camera optics. All of these are different for each LED color and unknown. This prevents us not only from putting an upper limit on the abundance of unresolved suspended particles, but also from calculating the surface reflectance for each LED color channel.

\subsection{Color and brightness variations}
\label{sec:variations}

\begin{figure}
\centering
\includegraphics[width=8cm]{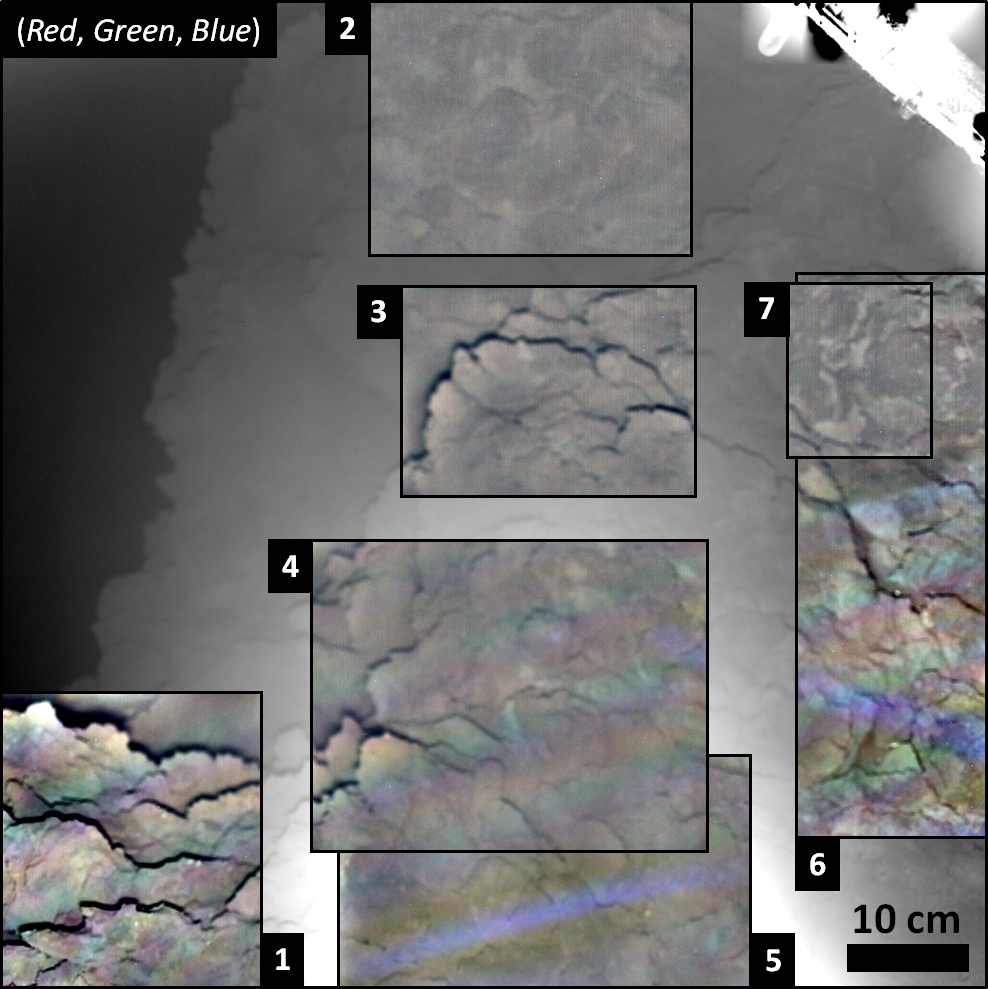}
\caption{Color composites were constructed for seven individual subframes from the {\it Red}, {\it Green}, and {\it Blue} images (\#4, \#2, \#1). They are shown superposed on the {\it Red} background image at their original location. The color of the subframes was ``corrected'' by dividing the images in the three color channels by individually fit polynomial surfaces. All composites are displayed with identical brightness scaling ($\pm 15$\% of the average). The scale bar is valid for the foreground.}
\label{fig:color}
\end{figure}

Because of the large amount of stray light and the large brightness gradient due to the LED illumination it is difficult to assess the degree of intrinsic color and brightness variations over the comet surface in the FOV. Stray light should be subtracted from the image and the illumination gradient should be divided out. But the distribution of both is different for each LED color and, moreover, unknown. We therefore turn to alternative techniques. First, we analyze the color variation in subframes of the full image frame. For each subframe we construct ``corrected'' RGB color composites from the {\it Red}, {\it Green}, and {\it Blue} images after dividing each of these by an individually fit fourth-order polynomial surface. Details on the polynomial surfaces and subframe definitions are provided in the supplementary material. Because of the comparatively small size of the subframes, this procedure better preserves relative brightness differences within the subframe than for the full frame. Figure~\ref{fig:color} displays the color subframes superposed on a background image for reference. Subframes in the bottom half of the full frame (1, 4, 5, 6) show the diagonal fringed color pattern (see Sec.~\ref{sec:overview}) that is absent from subframes in the top half of the frame (2, 3, 7). The transition is clearly visible in subframe 6, which extends over both the bottom and top half. The color fringes are optical artifacts. The broad particle trail in the {\it Blue} image shows up blue in subframes 5 and 6. The subframes are displayed in Fig.~\ref{fig:color} with the brightness scale of the individual color channels limited to $\pm 15$\% of the average value (unity). The color variations in the bottom part of the frame are of the order of 10\%. Subframes in the top part of the frame do not show appreciable color variation, which therefore must be smaller than 10\%. The surface in subframe~7 appears to show a bi-modal brightness distribution. We attempted to reconstruct the intrinsic color difference between two small adjacent patches, one relatively bright and the other dark, whose close proximity suggests that they experienced the same stray light and illumination regime. However, the amount of stray light proved difficult to estimate, as the gradients are large, even over this small subframe, and different for each LED color. The results are highly sensitive to the amount of stray light assumed, and thus we were not able to reliably establish the existence of a color difference between the patches of different brightness.

\begin{figure}
\centering
\includegraphics[width=\textwidth]{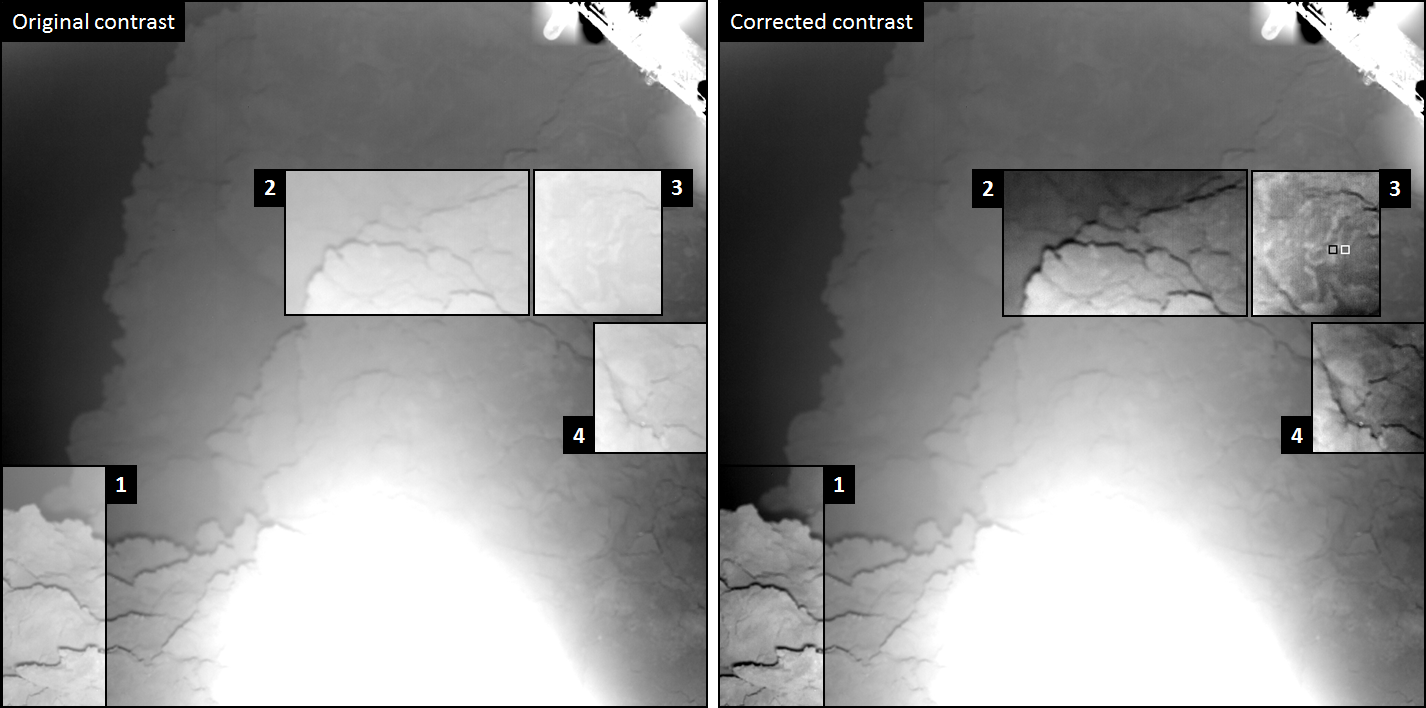}
\caption{The surface contrast was estimated for four individual subframes of the {\it Red} \#4 image by requiring shadows to be devoid of signal (black). The subframes are shown superposed on the {\it Red} background image (with log-scaled brightness for clarity) at their original location. {\bf Left}: Subframes shown at their original contrast (linear brightness scaling with black: zero signal, white: maximum subframe signal). {\bf Right}: Subframes shown at the corrected contrast. The small squares in subframe~3 were used for a contrast calculation (see text).}
\label{fig:contrast}
\end{figure}

We take a similar approach to estimate the brightness contrast intrinsic to the surface, concentrating on four subframes of {\it Red} image \#4. The original contrast in the subframes is low due to stray light (Fig.~\ref{fig:contrast}, left). To estimate the amount of stray light we assume that shadows are black. The small phase angle that results from the small distance between aperture and LED array leads to shadows, most clearly in the lower half of the frame. These are not perfectly black because of indirect illumination and infilling by the PSF of surrounding illuminated terrain, but especially the wider shadows (e.g.\ at bottom left) should be close to black. We force zero intensity in shadowed areas by subtracting a constant intensity from the entire subframe. This dramatically increases the contrast (Fig.~\ref{fig:contrast}, right). Some subframes now clearly show brightness variations intrinsic to the surface. For subframe~3, for example, we calculate the corrected, true contrast between neighboring dark and bright patches to be 57\% (meaning the bright patch is 57\% brighter than the dark patch). However, the shadows in this subframe may be too narrow in extent, and thereby not completely black due to infilling by the PSF associated with the surrounding illuminated terrain. If we assume that subframe~3 is affected by the same level of stray light as derived for neighboring subframe~2, the contrast is only 28\%, half the previous value.

\subsection{Surface morphology}
\label{sec:morphology}

\begin{figure}
\centering
\includegraphics[width=8cm]{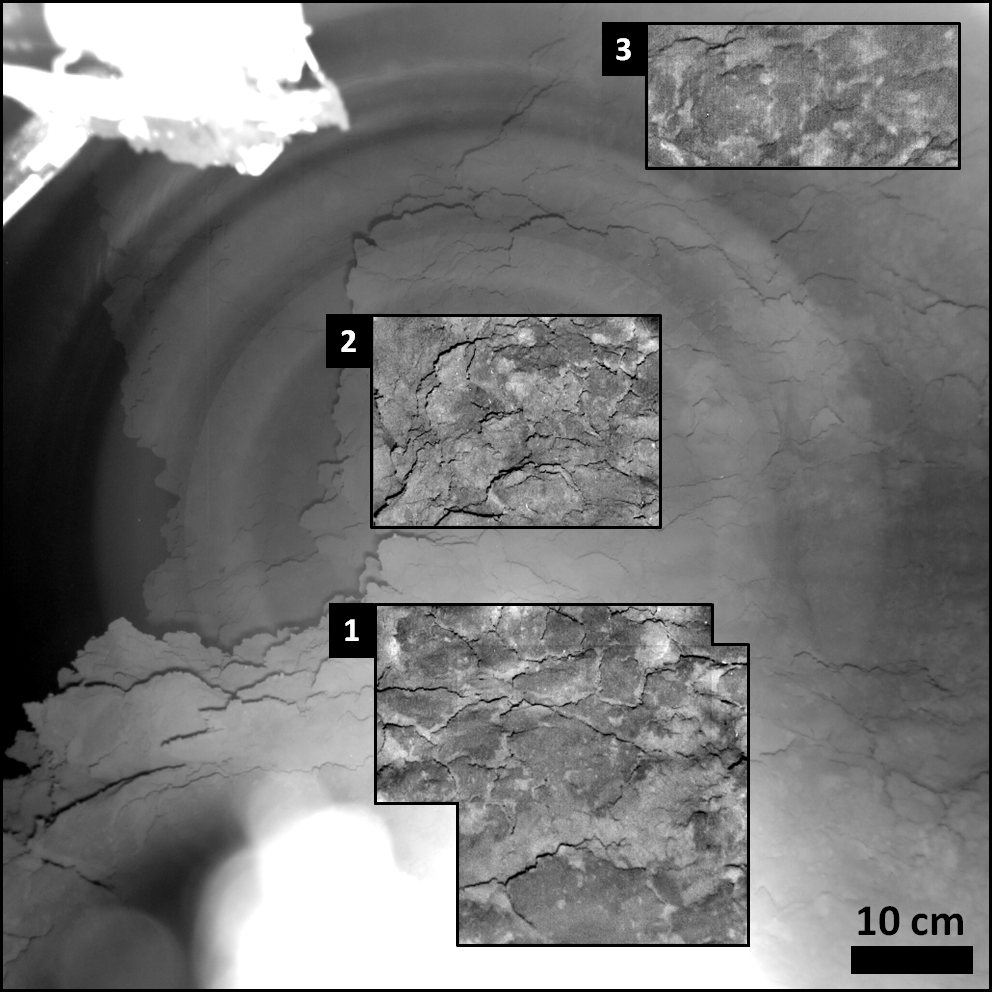}
\caption{Brightness variations in three subframes of the {\it Red} \#6 image shown at their original location (background image has log-scaled brightness). Each subframe has the original image divided by a best-fit polynomial surface. All subframes are displayed with identical (linear) brightness scaling (white and black are subframe mean $\pm 10$\%). Subframe~1 is a composite (average) of two individually corrected subframes. The scale bar is valid for the foreground.}
\label{fig:brightness_variation}
\end{figure}

The surface morphology can be best distinguished in the last ROLIS image, {\it Red} \#6, which was acquired after rotating and elevating the lander body slightly with respect to the landing position. As this image was taken at a larger distance to the surface than the earlier ones, the IFL was used to ensure that the surface would be fully in focus. In Fig.~\ref{fig:brightness_variation} we choose three subframes within the full image and divide them with a best-fit fourth-order polynomial surface to remove the large brightness gradients. Details on the polynomial surfaces and subframe definitions are provided in the supplementary material. Subframe~1 shows the surface at the highest spatial resolution achieved: 0.8~mm per pixel. The terrain in subframe~3 was more distant and was imaged at a resolution of about 1.0~mm per pixel. According to the DTM, the distance, and thereby resolution, of the terrain in subframe~2 was intermediate to that of the other subframes. The S/N in subframe~3 is lower than in the other subframes because of the large distance to the surface and high level of stray light in the image corner. The noise is dominated by photon noise. We calculated the average S/N for the signal coming from the surface in the three subframes in Table~\ref{tab:noise}. The expected S/N varies depending on the fraction of stray light (or light reflected off dust) one assumes. In Fig.~\ref{fig:stray_light} we demonstrated that, for one location at the edge of the image, only a third of the signal is coming directly from the surface (stray light fraction 0.65). Judging from Fig.~\ref{fig:brightness_variation}, the stray light fraction in this location appears to be relatively low, in other parts of the image, especially at the bottom of the image and in the corners, it must be higher. We therefore estimate the S/N in subframes~1 and 2 to be in the 30-60 range. The S/N in subframe~3 may be as low as 10.

\begin{figure}
\centering
\includegraphics[width=9cm]{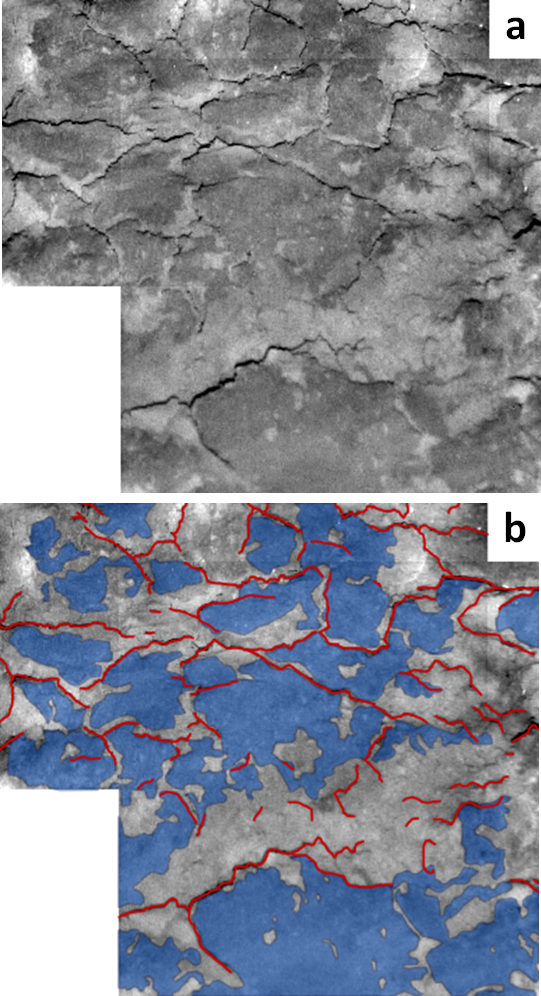}
\caption{Mapping the suspected plates visible in the ROLIS images. ({\bf a}) Subframe~1 from Fig.~\ref{fig:brightness_variation}. ({\bf b}) The red lines trace the narrow shadows in (a) and generally correspond to the plate edges. The relatively dark unit representing the plate surface is mapped in blue, whereas the relatively bright unit is left uncolored.}
\label{fig:geological_map}
\end{figure}

Recall that the DTM shows that the surface in subframes~1 and 3 is seen more or less face-on, i.e.\ at a relatively small angle ($<40^\circ$; see Fig.~\ref{fig:DTM}). The same may hold for the surface in the top right corner of subframe~2, but towards the left of the frame it slopes away steeply from the lander. The surface seen at a small angle displays a bimodal brightness distribution that appears to be intrinsic, with relatively dark areas that have a rather smooth appearance and relatively bright areas that appear more rough and clumpy. In subframe~1, the dark areas have the appearance of plates bounded by thin shadowed ridges. We mapped both the shadows and the dark areas in Fig.~\ref{fig:geological_map}. Especially in the top part of subframe~1 the narrow shadows appear to trace out the contours of plates\footnote{Because of the unusual illumination geometry, some readers might recognize the plate structure only when viewing Fig.~\ref{fig:geological_map}a upside down.}. The plate surface is mostly dark, while bright areas are concentrated at their edges. The dark/bright boundary often runs parallel to the narrow shadows (red lines in Fig.~\ref{fig:geological_map}b). In none of the subframes individual particles can be distinguished, meaning that any particles that make up this surface are very small ($\ll 1$~mm). We do not distinguish any pores either.

\begin{figure}
\centering
\includegraphics[width=8cm]{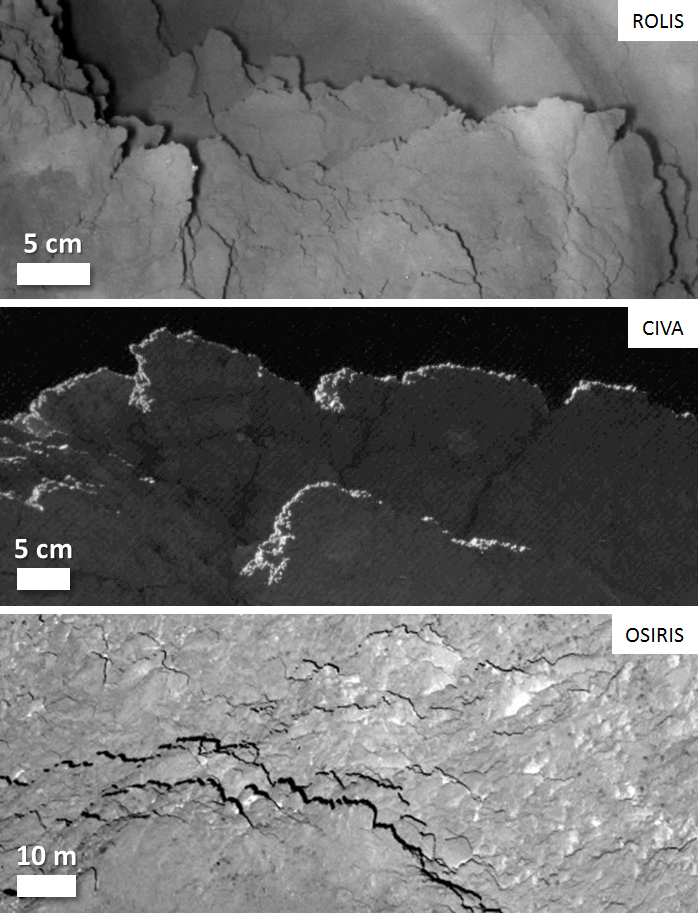}
\caption{The ridges visible in the ROLIS images ({\bf top}; detail from {\it Red} \#6) share morphological similarities with the back-lit cliff seen by CIVA camera~1 ({\bf middle}; \citealt{Bib15}) and rough terrain in a very low phase angle image taken by the Rosetta OSIRIS camera from 6.0~km altitude ({\bf bottom}). The ROLIS scale varies over the FOV; the scale bar assumes a 0.9~m distance to the surface. The CIVA scale is based on the 1~mm per pixel resolution estimate by \citet{P16}. OSIRIS image credits: ESA\slash Rosetta\slash MPS for OSIRIS Team MPS\slash UPD\slash LAM\slash IAA\slash SSO\slash INTA\slash UPM\slash DASP\slash IDA.}
\label{fig:comparison}
\end{figure}

There are morphological similarities between the scene observed by ROLIS and a back-illuminated cliff seen by the CIVA camera \citep{Bib15,P16} and comet terrain observed by OSIRIS at very low phase angle\footnote{\url{http://www.esa.int/spaceinimages/Images/2015/02/14_February_close_flyby}}. The scenes in Fig.~\ref{fig:comparison} represent different spatial scales, although the CIVA image scale is close to that of ROLIS. The appearance of the terrain is remarkably similar over all scales. The jagged contours are evidence for the fractal nature of the surface, which may somehow be related to cometary activity. Shadowed ridges are seen in the ROLIS and OSIRIS images, both of which were acquired at near zero phase angle. The OSIRIS scene has also been observed at larger phase angles and the terrain appears to be a slope of jumbled, rough material\footnote{\url{http://www.esa.int/spaceinimages/Images/2015/03/14_February_flyby_in_context_Osiris_and_NavCam}}. In the low phase angle image, the shadowed ridges create the appearance of layers, but they are in fact due to the presence of large, irregular outcrops. Smaller scale features, like the jumbled blocks and boulders, are rather subdued. Most likely, the ROLIS scene in Fig.~\ref{fig:comparison} represents just such a rough and sloping terrain, albeit on a much smaller scale. However, the plates with the bi-modal brightness distribution in Fig.~\ref{fig:geological_map} were seen more face-on, and we suspect that the shadowed ridges at their edges have a different explanation. Before making a definite interpretation we first investigate the consequences of imaging a surface at the ROLIS geometry by means of an experiment.

\section{Imaging experiment with textural analogs}
\label{sec:experiment}

To understand the implications of the bi-modal brightness distribution in Fig.~\ref{fig:geological_map} for the physical nature of the surface, we performed an experiment in which we imaged possible textural analogs for the comet surface with the flight spare camera under relevant illumination conditions (near zero phase). Rather than trying to create an accurate simulation of the comet surface under realistic environmental conditions, like the \citet{K98} and \citet{PP16} experiments, we assess whether textural variations might be responsible for the bi-modal brightness distribution. In particular, we evaluate the appearance of a low-albedo surface for two end-member cases for the consolidation state: (1) fine particles and (2) consolidated. We ensured that in both cases the surface possessed the same topography. As a material for the simulation we chose tiling grout powder, which is available in home improvement stores. The ``black'' variety we used has a low reflectance and consists mainly of cement powder, Ca-bearing minerals, and a coloring agent (see Appendix~\ref{sec:appendix} for a full characterization). This material is well suited to our purpose because it is particulate in its native, anhydrous form, and it solidifies into a crust when hydrated. The powder is very fine and individual particles are not resolved by the camera. The simulated surfaces were prepared in the following way. First, we distributed the powder over a large area ($1 \times 1$~m), sprinkling it by hand in order to create a rough, lumpy surface. To study the effects of roughness we subsequently created several $8 \times 8$~cm smooth impressions on this surface with the aid of an improvised plastic-covered stamp. The surface was prepared in three configurations: (a) soft powder, (b) hard crust, and (c) cracked crust. Surface~(b) was prepared from surface~(a) by hydration: We sprayed the surface with tap water in such a way that the water impacted from above as a fine mist of droplets, preserving the topography. It was then left to dry and harden overnight. Surface~(c) was prepared by slowly running surface~(b) over a small ridge, creating large parallel cracks.

It was not possible to put the experimental surface at the same distance as on the comet (1-2~m) because of the large size of the target required and the very low S/N. The latter was the inevitable result of the low surface reflectance and the high dark current of the CCD detector (the experiment was performed at room temperature). We increased the S/N of the images by subtracting the average of 10 dark exposures from the average of 10 illuminated exposures. The camera was placed in front of the surface in two positions: (1) boresight incidence angle $25^\circ \pm 5^\circ$, height above the surface $31.0 \pm 0.5$~cm, (2) incidence angle $60^\circ \pm 5^\circ$, height $24.0 \pm 0.5$~cm. The surface was in focus without using the IFL. Illumination was either provided by the camera LED array or the room lights. The phase angle in the LED images is small, typically $<5^\circ$. The room light provided ambient illumination resulting in a mixture of phase angles roughly in the $20^\circ$-$90^\circ$ range, i.e.\ far from zero.

\begin{figure}
\centering
\includegraphics[width=\textwidth]{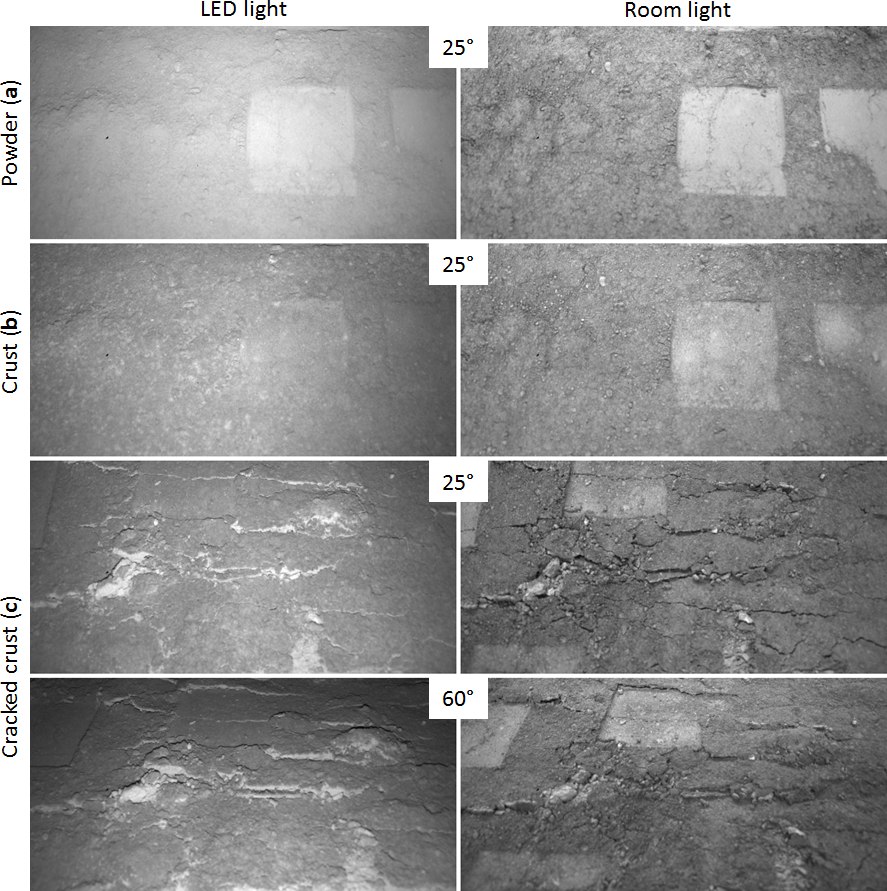}
\caption{Three types of artificial surfaces ({\bf a}, {\bf b}, {\bf c}) observed with the ROLIS flight spare camera using illumination by the {\it Red} LEDs ({\bf left}) or room lights ({\bf right}). Shown here is the lower half of the full image frame ($1024 \times 512$ pixels). The angle in the inset denotes the boresight incidence angle. The square imprints on the surface are sized $8 \times 8$~cm.}
\label{fig:experiment}
\end{figure}

Figure~\ref{fig:experiment} shows the surfaces as they appeared to the camera. The most striking aspect of the results is how different the surfaces appear depending on the type of lighting (LED or room light). We first discuss the LED illuminated surfaces in the left column of Fig.~\ref{fig:experiment}. The powder (a) has a rather soft appearance with the brightness mainly governed by the roughness of the terrain. The smooth squares stand out as relatively bright, and the lumpy texture of the surface can be clearly recognized. On the other hand, the crust (b) appears very different, now with the smooth squares blending in with the rough surroundings. The overall surface brightness is lower than that of the powder. Cracking the crust into plates (c) introduced ridges over the entire width of the FOV that cast narrow shadows. The shadows are more pronounced at higher incidence angles. The edges of the plates expose the original, unconsolidated powder below the crust, where the water did not penetrate. The powder is clearly visible as it is brighter than the overlying hard crust, creating a striking pattern of bright patches. As said, the surfaces appear very different in the ambient room lighting (right column of Fig.~\ref{fig:experiment}). The powder (a) and crust (b) surfaces now look similar in appearance and brightness, and the smooth squares are consistently  brighter, even for the crust. Cracking the crust (c) has introduced large-scale roughness that is recognized through shadowing. Surprisingly, the powder patches at the plate edges that stood out so clearly in the low-phase angle images are completely unremarkable. For small phase angles the dominant factor governing brightness is the aggregation state; light escapes more easily from the particulate surface. For phase angles far from zero this does not seem to be the case. As the phase angle in the ROLIS images is in the range of the opposition surge\footnote{The opposition surge is the non-linear increase of the surface reflectance towards zero phase angle when displayed on a logarithmic scale.}, we can attribute the brightness difference to either shadow hiding or coherent backscatter (e.g.\ \citealt{H98}). We stress that this brightness difference is related to the physical properties of the surface rather than resulting from mineralogical changes introduced by wetting.

Our experiment may offer clues to the nature of the comet surface. When we compare the results in Fig.~\ref{fig:experiment} with the comet image in Fig.~\ref{fig:geological_map} we notice similarities between the latter and the hard, cracked experimental surface. While the distance to surface was 3-4 times larger on the comet than during the experiments (resulting in smaller phase angles), the main observational results are reproduced: thin bands of shadows and dark plates with bright edges.

\section{Discussion}

The ROLIS images of the final landing site in the Abydos region reveal a surface that is fundamentally different from that imaged during the initial descent \citep{M15}. Whereas the first landing site was blanketed by a layer of centimeter to meter-sized debris, we cannot distinguish any particles at all in the post-landing images. The debris is thought to derive from ``airfall'', the deposition of particles originating from a different region on the comet \citep{T15}. As only the southern hemisphere experiences vigorous activity during perihelion, there is probably a net flux of particles from the southern to the northern hemisphere over the orbit around the Sun \citep{K15}. Both the first and final Philae landing site can therefore be regarded as a sink for airfall particles. Why we do not see these particles in ROLIS images of the final landing is not immediately clear. The OSIRIS images of Philae show the lander located below a large outcrop that could have shielded it from a ``particle wind''. They also appear to confirm that the surface below the lander (that is visible to ROLIS) is strongly inclined with respect to the local comet surface. In case deposition did take place, particles may have simply rolled off. However, the surface has a considerable roughness with some structures being oriented at right angles, and one may expect at least some centimeter-sized particles to be able to accumulate. In that case, airfall particles may have been removed, possibly by activity. The entire comet surface is probably active (e.g. \citealt{K15}), and \citet{L16} confirm Abydos as a region where erosion takes place, presumably driven by activity. The ROLIS close-up images may therefore provide insight into the nature of cometary activity, the mechanics of which are still not well understood \citep{G15}.

\begin{figure}
\centering
\includegraphics[width=11cm]{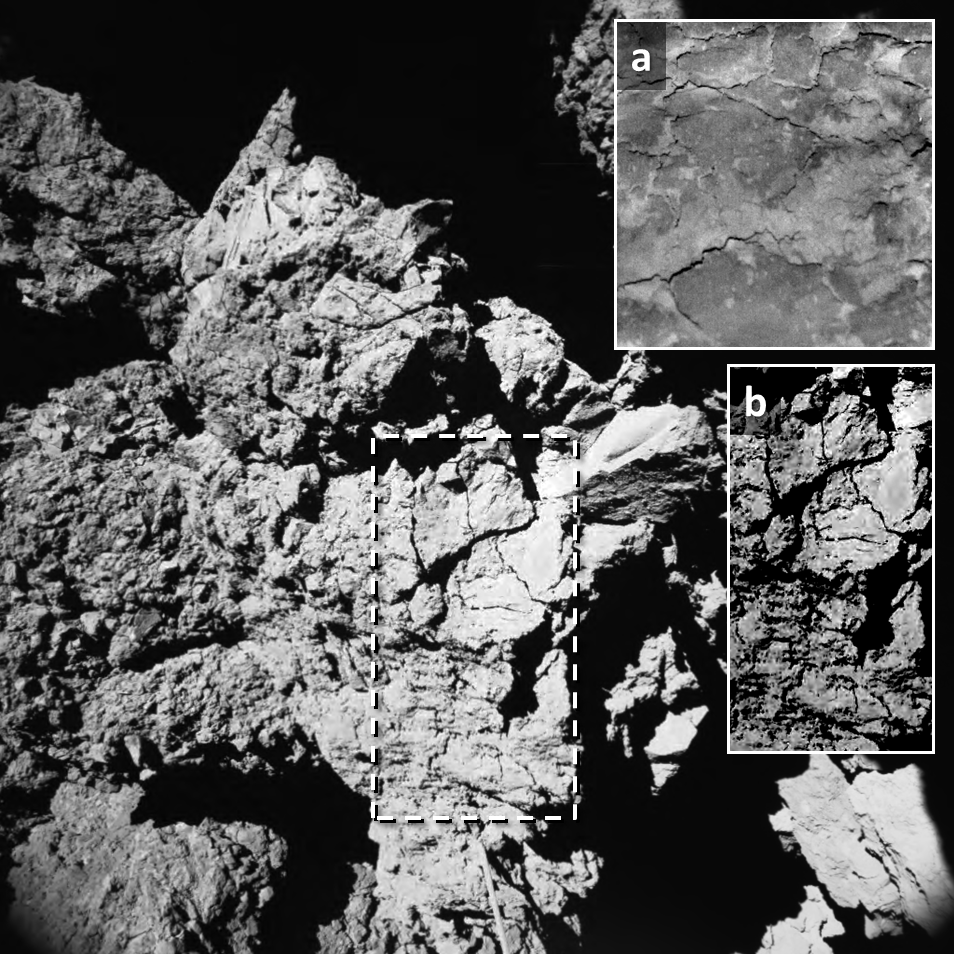}
\caption{The image acquired by CIVA camera~4 \citep{Bib15,P16}. The dashed line surrounds an area of cracked yet smooth surface that we consider most representative for terrain seen by ROLIS. Inset~({\bf a}) shows part of subframe~1 in Fig.~\ref{fig:brightness_variation} (ROLIS image). Inset~({\bf b}) shows the area within the dashed line with the contrast enhanced. Note that the enhancement also reveals compression artifacts that are concentrated in areas of large brightness contrasts.}
\label{fig:comparison_CIVA}
\end{figure}

ROLIS offers a unique perspective on the comet surface, not only because of the very high spatial resolution (0.8 mm per pixel) but also because of the low phase angles provided by the LED illumination. That is, the images allow assessment of variations in the strength of the opposition surge that result from the physical condition of the comet material. As such we can expect that the surface observed by ROLIS would look different from the perspective of the CIVA cameras, as they observed at phase angles far from zero \citep{Bib15}. The ROLIS scene itself was not observed by CIVA, as ROLIS looked downward and all CIVA cameras looked sideward. However, it is reasonable to assume, and the OSIRIS images of Philae confirm this (see Fig.~\ref{fig:Philae_found}), that ROLIS did not observe a unique type of terrain. It must therefore be possible to identify similar terrain in the CIVA images. We already pointed out morphological similarities with the top of the back-lit cliff seen by CIVA camera~1 (Fig.~\ref{fig:comparison}), but other CIVA cameras observed the surface well-illuminated by the Sun. The camera~4 image in Fig.~\ref{fig:comparison_CIVA} shows a rough outcrop next to an area of cracked plates that have a more smooth appearance \citep{P16}. We suggest that the latter terrain is representative for the plates in Fig.~\ref{fig:geological_map}, in which case the narrow shadows are associated with cracks between the plates. Cracks and fractures have been observed by OSIRIS all over the comet surface at scales from ten to hundreds of meters, and may predominantly result from thermal insolation weathering \citep{ElM15}. The ROLIS and CIVA images demonstrate that the surface is also cracked at smaller (decimeter) scales. The ROLIS plates are generally brighter towards the edges, whereas the CIVA plates do not appear as such (Fig.~\ref{fig:comparison_CIVA}, insets). The difference in phase angle may be responsible for this apparent contradiction. The \citet{K98} experiments suggest that cometary activity leads to the formation of a refractory dust mantle and a sub-surface layer of increased compressive strength, formed by recrystallization of water vapor. Could the relatively dark and bright surface units in Fig.~\ref{fig:geological_map} represent the dust mantle and the (exposed) ice-rich hard layer? This seems unlikely; while the bright unit appears to be widespread, exposed water ice would rapidly sublimate. Water ice was not identified on the pre-perihelion comet surface at a heliocentric distance of 3~AU \citep{CC15}. We therefore take clues from our experiment described in Sec.~\ref{sec:experiment}. A brightness difference would arise naturally if the plate surface is consolidated (smooth on a micro-scale) down to at least a millimeter, and the material towards the edges is unconsolidated (rough on a micro-scale).

One implication of this hypothesis might be that unconsolidated material is exposed by shedding of consolidated crust. A consolidated, refractory crust could be impermeable to water molecules coming from below, thereby allowing the gas pressure to increase until the layer is lifted off. Without a consolidated crust the decrease of permeability within the refractory layer on top of the sublimating ice is so small that the pressure due to sublimation can only remove refractory layers with extremely low tensile strength (e.g.\ \citealt{SB12,G15}). If the landing site was originally rich in airfall particles, their absence in the images might indicate that the plate surface is frequently shed in an ongoing process of activity. As to the nature of the consolidating material, we can only speculate. Perhaps organic materials play a role. \citet{Q16} infer the presence of a complex mixture of organic, semi-volatile material that is stable below 220~K, which is just above the temperature of the water sublimation front underneath the refractory cover \citep{K15}. This mixture may start to sublime at the higher surface temperatures reached in the inner Solar System \citep{K15}. The diurnal cycle of sublimation and re-condensation of organic material might consolidate the porous refractory matrix.

Results from the MUPUS instrument onboard Philae may support the existence of plates with a consolidated surface. Its infrared radiometer derived a relatively high thermal inertia, which was interpreted as evidence for a thinner layer of dust than average \citep{S15}. In light of the instrument's large FOV ($\pm 30^\circ$ around the line of sight), the relatively high value may be consistent with the presence of the hypothesized plates, with the consolidated plate surface being devoid of loose dust and its unconsolidated edges being dust-rich. The explanation in terms of the surface texture we offer for the bi-modal brightness variations observed by ROLIS at low phase angles is based on the assumption of compositional homogeneity and consistent with the morphological evidence. Other explanations are possible; there could be differences in composition and/or particle size between the two types of terrain that lead to a difference in strength of the opposition surge. To assess this possibility it would be necessary to perform photometric measurements of realistic surface analogs under comet conditions.

\section*{Acknowledgements}

The authors thank Wolfgang Bresch for his continued support during the development and operations of ROLIS and Stubbe Hviid for helpful suggestions.

%\appendix
\begin{appendices}

\section{Characterization of the experimental materials}
\label{sec:appendix}

\begin{figure}
\centering
\includegraphics[width=\textwidth]{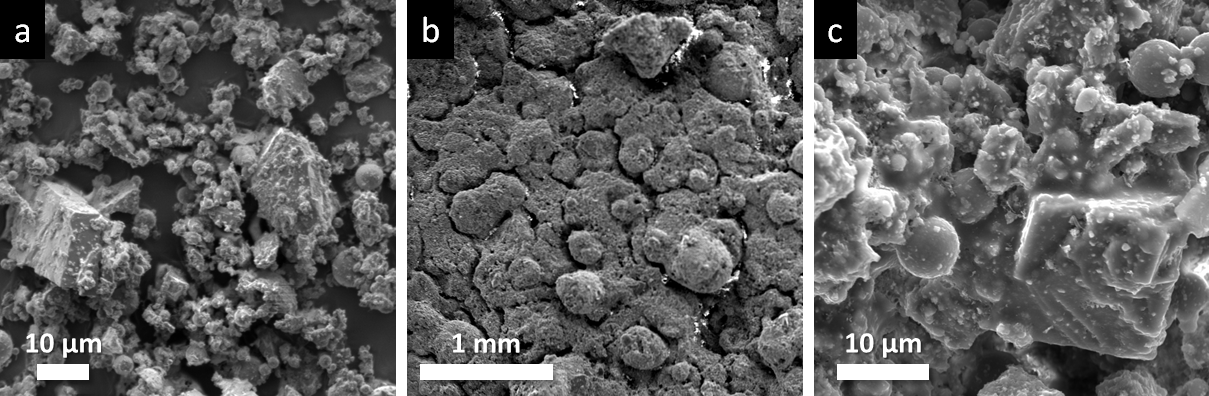}
\caption{Raster electron microscope images of the experimental surface materials. ({\bf a})~Powder, high magnification. ({\bf b})~Crust, low magnification. ({\bf c})~Crust, high magnification.}
\label{fig:REM}
\end{figure}

The physical and chemical properties of the powder and crust surfaces used in the experiment (Sec.~\ref{sec:experiment}) were analyzed by the {\it Gesellschaft zur F\"orderung der naturwissenschaftlich-technischen Forschung} (GNF) in Berlin, Germany. The powder is composed of (1) spherical particles that mostly consist of \ce{SiO2} and may also contain \ce{Al2O3}, (2) Ca-rich particles, which were mostly produced by crushing larger grains, (3) small particles that originate from abrasion of the previous particle types and Fe-rich clays, and (4) agglomerates of the small particles produced by sintering. The particle size distribution was measured by means of a laser with the particles immersed in a liquid, either water or ethanol. The majority of the particles has a size far below 100~\textmu m, with an upper limit of about 200~\textmu m. The most frequent particle size is 10~\textmu m (both liquids give the same results), but this size is probably that of agglomerates of even smaller particles. Raster electron microscope (REM) images of the powder (Fig.~\ref{fig:REM}a) suggest the actual average particle size to be 2-5~\textmu m. Wetting the powder leads to the formation of a \ce{Ca2SiO4} gel that fills the space between the particles and which hardens into a crust as the surface dries. The REM images of the crust at low (Fig.~\ref{fig:REM}b) and high (Fig.~\ref{fig:REM}c) resolution show how this in-filling of the inter-particle space produces a smooth surface structure. The introduction of smoothness and the filling in of gaps at the micrometer scale alters the reflective properties of the powder surface in a fundamental way.

\end{appendices}

%\end{linenumbers}

\bibliography{rolis}

\newpage
\clearpage

\begin{table}
\centering
\caption{Details of all post-landing images acquired by ROLIS. The file names refer to the raw images as archived. Calibrated images have `FS3' in the file name instead of `FS2'. Images 1-5 were part of a pre-programmed set, and image 6 was acquired {\it ad hoc}. The {\it Dark} image had the LEDs switched off. Time is UTC at the start of the exposure. The last column refers to the position of the infinity lens (IFL) with respect to the optical path. With the IFL out the optical path the camera was focused at an object distance of 30~cm.}
\vspace{5mm}
\begin{tabular}{lllllrl}
\hline
\# & File name & Date & Time       & LED        & $t_{\rm exp}$ (s) & IFL \\
\hline
1 & ROL\_FS2\_141113001043\_336\_01 & 2014-11-13 & 00:10:43 & {\it Blue}  & 30.000 & Out \\
2 & ROL\_FS2\_141113001304\_336\_02 & 2014-11-13 & 00:13:04 & {\it Green} & 25.606 & Out \\
3 & ROL\_FS2\_141113001333\_336\_03 & 2014-11-13 & 00:13:33 & {\it Dark}  & 1.536  & Out \\
4 & ROL\_FS2\_141113001408\_336\_04 & 2014-11-13 & 00:14:08 & {\it Red}   & 5.763  & Out \\
5 & ROL\_FS2\_141113001504\_336\_05 & 2014-11-13 & 00:15:04 & {\it IR}    & 6.157  & Out \\
6 & ROL\_FS2\_141114232010\_336\_00 & 2014-11-14 & 23:20:10 & {\it Red}   & 5.763  & In \\
\hline
\end{tabular}
\label{tab:images}
\end{table}

\begin{table}
\centering
\caption{The average signal and noise (in DN) for the three subframes in Fig.~\ref{fig:brightness_variation} (image \#6, bias-corrected). The four rightmost columns list the average S/N for the signal coming from the surface assuming different fractions of stray light (in brackets). We adopted 2~DN for the combined system and read-out noise \citep{M07}.}
\vspace{5mm}
\begin{tabular}{lllrrrr}
\hline
\# & Signal & Noise & S/N (0) & (0.3) & (0.6) & (0.9) \\
\hline
1 & $3298 \pm 807$ & $19.2 \pm 2.1$ & 172 & 120 & 69 & 17 \\
2 & $1742 \pm 155$ & $14.6 \pm 0.6$ & 119 & 84  & 48 & 12 \\
3 & $1172 \pm 50$  & $12.3 \pm 0.2$ & 95  & 67  & 38 & 10 \\
\hline
\end{tabular}
\label{tab:noise}
\end{table}

\newpage
\clearpage

%\newpage
%\clearpage

\begin{figure}
\centering
\includegraphics[width=\textwidth]{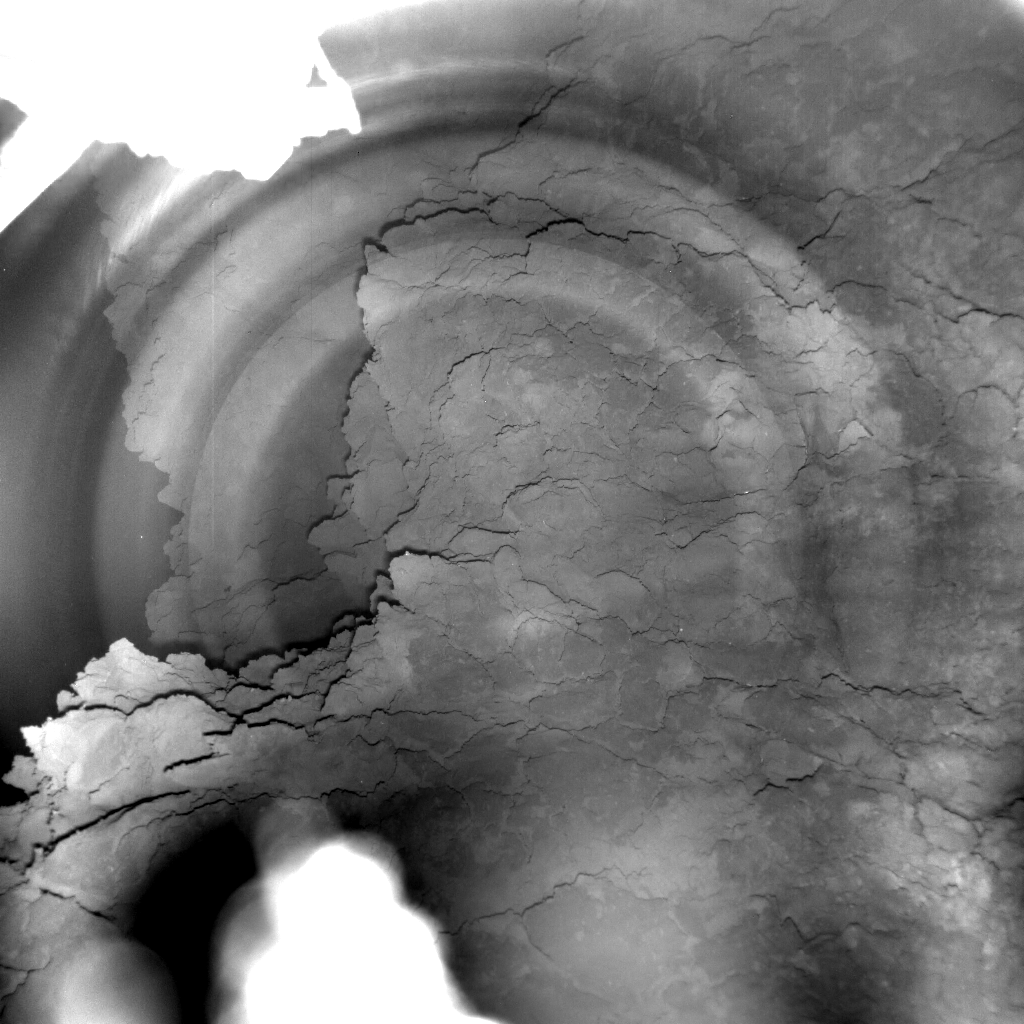}
\caption{The last image acquired by ROLIS on the comet surface ({\it Red} \#6), enhanced by division by a best-fit polynomial surface.}
\label{fig:last_red}
\end{figure}

%\newpage
%\clearpage

%\newpage
%\clearpage

%\newpage
%\clearpage

%\newpage
%\clearpage

%\newpage
%\clearpage

%\newpage
%\clearpage

%\newpage
%\clearpage

%\newpage
%\clearpage

%\newpage
%\clearpage

%\newpage
%\clearpage

%\newpage
%\clearpage

%\newpage
%\clearpage

%\newpage
%\clearpage

%\newpage
%\clearpage

\end{document}